\documentclass[11pt]{article}  
\usepackage{epsfig,graphicx,graphics,amsmath}  
\input epsf  
  
\newcommand{\beq}{\begin{equation}}  
\newcommand{\eeq}{\end{equation}}  
\newcommand{\ben}{\begin{enumerate}}
\newcommand{\een}{\end{enumerate}}
\newcommand{\bitem}{\begin{itemize}}  
\newcommand{\eitem}{\end{itemize}}
\newcommand{\bfig}{\begin{figure}}
\newcommand{\efig}{\end{figure}}
\newcommand{\bcen}{\begin{center}}
\newcommand{\ecen}{\end{center}}

\newcommand{\delete}[1]{}

\title{Galactic Rotation Described with Thin-Disk Gravitational Model}
\author{James Q. Feng and C. F. Gallo}

\begin{document}

\maketitle

\bcen
\Large{\bf Abstract}
\ecen


The measured rotation velocity profiles of mature spiral galaxies are successfully described with a gravitational model consisting of a thin axisymmetric disk of finte radius. The disk is assumed uniformly thin but with variable radial mass density. The governing integral equation is based on mechanical balance between Newtonian gravitational and centrifugal forces (due to galaxy rotation) at each and every point in a finite set of concentric rings. The nondimensionalized mathematical system contains a dimensionless parameter we call ``galactic rotation parameter'' which concisely crystallizes perspective. Computational solutions are obtained for the radial mass distributions that satisfy the measured rotational velocity profiles. Together with a constraint equation for mass conservation, the galactic rotational parameter is also determined from which the total galactic mass is calculated from measured galactic radii and maximum rotation velocities. These calculated total galactic masses are in good agreement with data. Our deduced exponentially decreasing mass distributions in the central galactic core are in agreement with almost all others. However our mass distributions differ toward the galactic periphery with more ordinary baryonic mass in these outer disk regions which are cooler with lower opactiy/emissivity (and thus darker).    


\section{Introduction} 
   
\subsection{Galactic Rotational Velocity Profiles from Observational Data}
The data on galactic rotational velocity profiles 
(Refs.\cite{Rubin1}-\cite{deBlockMH}) of mature spiral galaxies may be idealized as  
\beq \label{eq:smooth-V}
V(r) = 1 - e^{-r / R_c} \, , 
\eeq
where $V(r)$ denotes the tangetial velocity and 
$r$ the radial coordinate from the galactic center.
The parameter $R_c$ is a description of the various radii of the "cores" of different galaxies. Typical galactic rotational profiles described by (\ref{eq:smooth-V})
are displayed in Fig 1. As indicated by the measurement data, 
the rotation velocity typically rises linearly from the galactic center 
(as if the local mass was in rigid body rotation), and then reach an approximately constant (flat) velocity out to the galactic periphery.

\subsection{Historical Background: Keplerian Dynamics and Dark Matter} 

The observed galactic rotation curves described above and in Fig.1 are very different from the
Keplerian rotation of our solar-planet system in which the orbital velocity decreases inversely with square root of the radial postion. 
This Keplerian rotation comes from the {\em orbital velocity law} $M_r = r \, V^2(r) / G$, where $G$ is the gravitational constant and $M_r$ is the integrated mass enclosed within a spherical surface of $r$ (called orbital distance).    
This orbital velocity law is based on a simple spherically symmetric gravitational field (e.g., from a dominant central solar point mass). 
It was often applied to the dynamics of a galaxy system, where the 
mass appears to be distributed without the obvious spherical symmetry,   
to calculate the galactic mass distribution from the measured galactic rotation curves  (Ref.{\cite{Bennett}).  
Such a practice led to the conclusion that the galactic mass {\em must increase} with radial position. 
     
By contrast, the measured galactic luminosity curves decrease exponentially from the galactic center. 
Employing the concept that mass and light are directly relate via mass-to-light ratio, the exact opposite conclusion is reached (Ref.{\cite{Bennett}) that the mass {\em decreases} with radial position. 

It is this sharp discrepancy between these two opposite conclusions that has led to the concept of Dark Matter (Ref.{\cite{Bennett}), which inspired considerable research efforts in modern cosmology. To summarize this situation, ordinary baryonic matter decreases exponentially from the galactic center (as deduced from measured light profiles), whereas Dark Matter increases from the galactic center to the periphery to explain the rotational profiles (Eq.1 and Fig.1).       

Different aspects of the applicability of Keplerian dynamics will be discussed in more detail in \S~7.

\subsection{Prior Gravitational Disk Models with Assumed Mass/Light Ratio}

The internal gravitational aspects of galactic disks are very different than Keplerian (Ref.\cite{BT}), depending upon mass distribution and the interior locations under investigation. 
In most previous research employing appropriate gravitational disk geometry (Refs.\cite{BT}-\cite{Kruit}), it is assumed that the galactic density decreases exponentially with radius analogous to the measured light distribution (Refs.\cite{deJong1}-\cite{deJong2}). With this assumption, these prior models do not describe the measured velocity profiles, and recourse is again made to Dark Matter or gravitational deviations (Ref.\cite{Milogram}) to compensate.  More specifically, the Dark Matter is assumed to be distributed in massive peripheral spherical halos around the galaxies.

\subsection{Prior Gravitational Disk Models with Mass Distributions Determined by Computation from the Measured Rotation Profiles}

Again note the interior gravitational aspects of galactic disks is very much different than Keplerian (Ref.\cite{BT}). Other previous research employing disk models (Refs.\cite{Marmet}-\cite{Lusanna}) avoid assumptions re the unknown galactic density and computationally solve for mass distributions that satisfy the measured rotational velocity profiles. They assume only Newtonian gravity/dynamics or General Relativity. They successfully describe the measured rotation velocity profiles of mature spiral galaxies without any need for gravity deviations or massive peripheral spherical halos of Dark Matter. All the deduced mass distributions are perfectly reasonable and rational. The details utilized in these numerous studies vary, but the overall computational approach, guided by verified physical laws (Newtonian gravity/dynamics or General Relativity), is the common feature that yields successful description of the measured galactic rotational velocity profiles. Our approach is in this vein.

\subsection{Thin Disk Model with Mass Distribution Determined by Gravitational Computation}

Attention is focused on the internal gravitational aspects of galactic disks as influenced by the radial mass distribution (Ref.\cite{BT}). This disk model avoids assumptions re the unknown galactic density and computationally solves for mass distributions $\rho$ that satisfy the measured rotational velocity profiles. Only Newtonian gravity/dynamics is assumed. 
This model utilizes a finite axisymmetric disk of uniform thickness but variable radial density. 
The results successfully describe the measured rotation velocity profiles $V(r)$ of mature spiral galaxies (Refs.\cite{Rubin1}-\cite{deBlockMH}) without any need for gravity deviations or massive peripheral spherical halos of Dark Matter. All the deduced mass distributions are perfectly reasonable and rational. Prior modeling efforts have struggled to describe the rapid linear velocity rise in the galactic ``core'' and simultaneously describe the constant rotational velocity beyond the core and to the periphery. Our efforts have succeeded. Ordinary baryonic matter is found within the galactic disk but distributed more towards the galactic periphery which is cooler with lower opacity/emissivity (and therefore darker). Our total galactic mass determinations are also in good agreement with data. There are no mysteries in our approach or conclusions.

\section{Mathematical Formulation and Computational Techniques}

Our model consists of a finite axisymmetric thin disk of uniform thickness ($h$) but variable radial density ($\rho$) . Newtonian gravity and Newtonian mechanics are assumed. The gravitational forces are balanced against the centrifugal forces at each and every point. The goal is to determine a physically meaningful radial mass density distribution within the disk to satisfy a given target rotational velocity profile.

\subsection{Governing Equations} 

In the steady state, the gravitational forces are balanced against the centrifugal forces at each and every point in the finite series of concentric rings as follows. 
\beq \label{eq:force-balance0}
\int_0^1 \left[\int_0^{2 \pi} 
\frac{(\hat{r} \cos \phi - r) d\phi}
{(\hat{r}^2 + r^2 - 2 \hat{r} r \cos \phi)^{3/2}}\right] 
\rho(\hat{r}) h \hat{r} d\hat{r}
+ A \frac{V(r)^2}{r}
 = 0 \, ,
\eeq
where all the variables are made dimensionless by measuring lengths (ex., $r$, $\hat{r}$, $h$) 
in units of the outermost galactic radius $R_g$, 
disk mass density ($\rho$) in units of
$M_g / R_g^3$ with $M_g$ denoting the total galactic mass, 
and velocities [$V(r)$] in units of the maximum target galactic rotational velocity $V_{max}$.  
The disk thickness $h$ is assumed to be constant and small in comparison with the galactic radius $R_g$. Our results are insensitive to the exact value of this ratio as long as it is small.  
The present model determines the effective surface mass density on the disk,
i.e., the combined variable $(\rho \, h)$. The gravitational forces of the finite series of concentric rings is described by the first term (double integral) while the centrifugal forces are described by the second term. 

We have extracted/consolidated the relevant variables into a convenient transparent dimensionless ``galactic rotation parameter'' $A$ given by
\beq \label{eq:parameter-A}
A \equiv \frac{V_{max}^2 \, R_g}{M_g \, G} \, ,
\eeq
where $G$ denotes the gravitational constant, $R_g$ is the outermost galactic radius, and $V_{max}$ is the maximum asymptotic rotational velocity. This convenient parameter $A$ schematically displays the balance between the gravitational and centrifugal forces. 
For typical galactic values of $R_g$, $V_{max}$, and $M_g$ we have $A \sim 1.5$ as will be discussed in detail later. 


The total mass of the galaxy $M_g$ is determined by the constraint   
\beq \label{eq:mass-conservation}
2 \pi \int_0^1 \rho(\hat{r}) h \hat{r} d\hat{r} = 1.  
\eeq

The integral with respect to $\phi$ in (\ref{eq:force-balance0}) can be written as 
\beq \label{eq:elliptic-integral-form}
\int_0^{2 \pi} 
\frac{(\hat{r} \cos \phi - r) d\phi}
{(\hat{r}^2 + r^2 - 2 \hat{r} r \cos \phi)^{3/2}}
 = 2 \left[\frac{E(m)}{r (\hat{r} - r)} - \frac{K(m)}{r (\hat{r} + r)}\right]
 \, ,
\eeq
where $K(m)$ and $E(m)$ denote the complete elliptic integrals of the first kind and second kind,
with 
\beq \label{eq:m-def}
m \equiv \frac{4 \hat{r} r}{(\hat{r} + r)^2} \, .
\eeq
Thus, (\ref{eq:force-balance0}) becomes
\beq \label{eq:force-balance}
\int_0^1 \left[
\frac{E(m)}{\hat{r} - r} - \frac{K(m)}{\hat{r} + r}
\right] 
\rho(\hat{r}) h \hat{r} d\hat{r}  
+ \frac12 A V(r)^2
 = 0 \, .
\eeq
Equations (\ref{eq:force-balance}) and (\ref{eq:mass-conservation}) are used to determine the mass density distribution $\rho(r)$ in the disk, the galactic rotation parameter $A$, and the total galactic mass $M_g$, all from measured values of $V(r)$, $R_c$, $R_g$ and $V_{max}$. This is a completely defined problem deducible from the input data. 

NOTE: Some previous researchers (Ref.\cite{BT}) take the disk-appropriate elliptic integrals from the galactic center to infinity. We believe this approach is contrary to reality and tends to make the rotational velocity incorrectly decrease towards the galactic edge. By contrast, our research assumes that beyond the "galactic-rim-edge" the mass density drops quickly to the INTER-galactic level which is roughly spherically symmetric (not disk-shaped) and thus does not affect the galactic rotation.

\subsection{Discretization}
The governing equations (\ref{eq:force-balance}) and 
(\ref{eq:mass-conservation}) 
can be discretized by dividing the one-dimensional 
problem domain $[0, 1]$ into a finite number of line segments
called (linear) elements.
Each element covers a subdomain confined by two end nodes,
e.g., element $i$ corresponds to the subdomain
$[r_i, r_{i+1}]$, where $r_i$ and $r_{r+1}$ are nodal values of
$r$ at nodes $i$ and $i+1$, respectively.
On each element, which is mapped onto a unit line segment $[0, 1]$ in
the $\xi$-domain (i.e., the computational domain), the unknown
$\rho$ is expressed in terms of the linear basis functions as
\beq \label{eq:rho-xi}
\rho(\xi) = \rho_i (1 - \xi) + \rho_{i+1} \xi \, , \quad 0 \le \xi \le 1 \, ,
\eeq
where $\rho_i$ and $\rho_{i+1}$ are nodal values of $\rho$ at
nodes $i$ and $i + 1$, respectively.
Similarly, the radial coordinate $r$ on each element is also expressed
in terms of the linear basis functions by 
so-called isoparameteric mapping:
\beq \label{eq:r-xi}
r(\xi) = r_i (1 - \xi) + r_{i+1} \xi \, , \quad 0 \le \xi \le 1 \, .
\eeq
The $N$ nodal unknowns of $\rho_i = \rho(r_i)$ are determined by 
solving $N$ independent residual equations over $N - 1$ element obtained from
the collocation procedure, i.e.,
\beq \label{eq:force-balance-residual}
\sum_{n = 1}^{N - 1} \int_0^1 \left[
\frac{E(m_i)}{\hat{r}(\xi) - r_i} - \frac{K(m_i)}{\hat{r}(\xi) + r_i}
\right] 
\rho(\xi) h \hat{r}(\xi) \frac{d\hat{r}}{d\xi} d\xi
+ \frac12 A V(r_i)^2
 = 0 \, ,
\eeq
with
\beq \label{eq:mi-def}
m_i(\xi) \equiv \frac{4 \hat{r}(\xi) r_i}{[\hat{r}(\xi) + r_i]^2} \, .
\eeq
The values of $A$ and $M_g$ are solved by the addition of 
the constraint equation
\beq \label{eq:mass-conservation-residual}
2 \pi \sum_{n = 1}^{N - 1} \int_0^1 
\rho(\xi) h \hat{r}(\xi) \frac{d\hat{r}}{d\xi} d\xi - 1 = 0 \, .
\eeq
Thus, we have $N + 1$ independent equations for determining 
$N + 1$ unknowns. The mathematical problem is well-posed and completely determined.  

With Rc in (1) as an adjustable parameter, linear equations (10) and (12) 
for N + 1 unknowns can be solved with a standard matrix solver, e.g.,
by Gauss elimination (Ref.\cite{PressTVF}).

\subsection{Treatments of Singular Elements}
The complete elliptic integrals of the first kind and second kind can 
be numerically computed with the formulas (Ref.\cite{AbramowitzStegun}) 
\beq \label{eq:K-m1}
K(m) = \sum_{l = 0}^4 a_l m_1^l - \log(m_1) \sum_{l = 0}^4 b_l m_1^l 
\eeq
and
\beq \label{eq:E-m1}
E(m) = 1 + \sum_{l = 1}^4 c_l m_1^l - \log(m_1) \sum_{l = 1}^4 d_l m_1^l \, ,
\eeq
where
\beq \label{eq:m1-def}
m_1 \equiv 1 - m = \left(\frac{\hat{r} - r}{\hat{r} + r}\right)^2 \, .
\eeq
Thus, the terms associated with $K(m_i)$ and $E(m_i)$ in 
(\ref{eq:force-balance-residual}) become singular when $\hat{r} \to r_i$ 
on the elements with $r_i$ as one of their end points.

The logarithmic singularity is treated by converting the 
singular one-dimensional integrals into non-singular two-dimensional integrals
by virtue of the identities (Ref. \cite{Feng}): 
\beq \label{eq:log-integral-identities}
\left\{
\begin{split}
\int_0^1 f(\xi) \log \xi d\xi = - \int_0^1 \int_0^1 f(\xi \eta) d\eta d\xi 
\quad \quad \quad
\\  
\int_0^1 f(\xi) \log(1 - \xi) d\xi = 
- \int_0^1 \int_0^1 f(1 - \xi \eta) d\eta d\xi  \,
\end{split}
\right . \, ,
\eeq
where $f(\xi)$ denotes a well-behaving (non-singular) function of $\xi$
on $0 \le \xi \le 1$.

But a more serious non-integrable
singularity $1 / (\hat{r} - r_i)$ exists due to
the term $E(m_i) / (\hat{r} - r_i)$ in 
(\ref{eq:force-balance-residual}) as $\hat{r} \to r_i$. 
The $1 / (\hat{r} - r_i)$ type of singularity is treated by
taking the Cauchy principle value to obtain meaningful evaluation. 
In view of the fact that each $r_i$ is considered to be shared by two
adjacent elements covering the intervals $[r_{i-1}, r_i]$ and 
$[r_i, r_{i+1}]$, the Cauchy principle value of 
the integral over these two elements is given by 
\beq \label{eq:CPV-def}
\lim_{\epsilon \to 0} \left[
\int_{r_{i-1}}^{r_i(1 -\epsilon)} \frac{\rho(\hat{r}) \hat{r} d\hat{r}}{
\hat{r} - r_i} 
+ \int_{r_i(1 + \epsilon)}^{r_{i+1}} \frac{\rho(\hat{r}) \hat{r} d\hat{r}}{
\hat{r} - r_i}\right]
\, .
\eeq
In terms of elemental $\xi$, (\ref{eq:CPV-def}) is equivalent to 
\begin{eqnarray} \label{eq:CPV-xi}
-\lim_{\epsilon \to 0} \left[
\int_0^{1 -\epsilon} \frac{[\rho_{i-1} (1 - \xi) + \rho_i \xi] 
[r_{i-1} (1 - \xi) + r_i \xi] d\xi}{ 
1 - \xi} 
\hspace*{0.5in} \right .
\nonumber \\ \left . 
-\int_{\epsilon}^1 \frac{[\rho_{i} (1 - \xi) + \rho_{i+1} \xi] 
[r_i (1 - \xi) + r_{i+1} \xi] d\xi}{ 
\xi} \right]
\, .
\end{eqnarray}
Performing integration by parts on (\ref{eq:CPV-xi}) yields 
\begin{eqnarray} \label{eq:CPV-xi+}
-\left[
\int_0^1 \frac{d\{[\rho_{i-1} (1 - \xi) + \rho_i \xi] 
[r_{i-1} (1 - \xi) + r_i \xi]\}}{d\xi} \log(1 - \xi) d\xi
\hspace*{0.5in} \right .
\\ \left . 
+\int_0^1 \frac{d\{[\rho_{i} (1 - \xi) + \rho_{i+1} \xi] 
[r_i (1 - \xi) + r_{i+1} \xi]\}}{d\xi} \log \xi d\xi
\right]
\, ,
\end{eqnarray}
where all the terms associated with $\log \epsilon$ are either 
cancelling out each other or approaching zero at the limit of
$\epsilon \to 0$.

At the galaxy center $r_i = 0$,
\[
\int_{r_i}^{r_{i + 1}} \frac{\rho(\hat{r}) \hat{r} d\hat{r}}{
\hat{r} - r_i} = \int_0^{r_{i + 1}} \rho(\hat{r}) d\hat{r} \, . 
\]
Thus, the $1/(\hat{r} - r_i)$ type of singularity disappears naturally.

When $r_i = 1$, it is the end node of domain.  
We can either impose a boundary condition for $\rho(1)$ 
or still imagine another element extending beyond the domain boundary covering an interval 
$[r_i, r_{i+1}]$, because it is needed to provide useful 
Cauchy principle value.  However, this extra element can be assumed to 
cover a diminishing physical space, namely, $r_{i+1} \to r_i$ 
and $\rho_{i+1} \to \rho_i$ 
(in this extra element) such that its existence becomes numerically inconsequential.
Thus, we can safely have 
\[
\begin{split}
\int_0^1 \frac{d\{[\rho_{i} (1 - \xi) + \rho_{i+1} \xi] 
[r_i (1 - \xi) + r_{i+1} \xi]\}}{d\xi} \log \xi d\xi  
\quad \quad \quad \quad \quad \quad
\\
=  (\rho_{i + 1} - \rho_i) \int_0^1 r(\xi) \log \xi d\xi 
+ (r_{i+1} - r_i) \int_0^1 \rho(\xi) \log \xi d\xi  \to 0 \, .
\end{split}
\]
Now that only logarithmic singularities 
are left, (\ref{eq:log-integral-identities}) can be used to eliminate 
all singularities in integral computations.

In addition, to avoid cusps in mass density at the galactic center, 
continuity of the derivative of $\rho$ at the galaxy center $r = 0$ is applied. 
This boundary condition is imposed at the first node $i = 1$
to require $d\rho / dr = 0$ at $r = 0$. In discretized form
\beq \label{eq:rho-1}
\rho(r_1) = \rho(r_2)
\, .
\eeq


\section{Computational Results for Galactic Mass Distributions}
The computed galactic mass distributions that satisfy the target galactic rotational velocity curves (Eq 1 and Fig 1) (idealized from measurements) are shown in Fig 2. 
From the galactic center, the mass density tends to decrease steeply. However, beyond $R_c$, the mass density distributions decrease more slowly towards the galactic periphery.  This is reasonable since both the temperature and opacity/emissivity are lower towards the periphery and we would expect the light distribution to decrease more quickly towards the periphery.
The mass density distribution does not look like a simple exponential function, as that
deduced from the luminosity data.
This is the fundamental inaccuracy of the assumption (invoked by others) that the mass distribution follows the light distribution exactly. Our computational technique avoids that inaccurate assumption and yields rational reasonable mass distributions that yield agreement with the measured galactic rotational profiles.

\section{Ordinary Baryonic Matter versus Dark Matter}
To theoretically describe the measured rotational velocity curves of spiral galaxies, there
are two very different approaches and conclusions. 

(1) Ordinary Baryonic Matter. 
We assume Newtonian gravity/dynamics and computationally solve for successful mass distributions that duplicate the measured rotational velocities. These mass distributions decrease roughly exponentially from the galactic center, but then decrease more slowly (inversely with radius) towards the periphery. This decrease is slower than the measured light distribution. Thus there is ordinary baryonic matter within the galactic disk distributed towards the cooler periphery with lower emissivity/opacity and therefore darker. There are no mysteries in this rational scenario based on verified physics.   

(2) Dark Matter.  
By contrast, others inaccurately assume the galactic mass distributions follow the measured light distributions (approximately exponential), and then the measured rotational velocity curves are not  duplicated. But this assumption of a simple direct relationship between light intensity and mass is very inaccurate.  This so-called Mass/Light ratio is inaccurate since both the temperature and opacity/emissivity are important but ignored variables. These deficiencies are clear from edge-on views of spiral galaxies where a dark galactic line is obvious against a bright galactic background. revealing the substantial radial temperature gradient across the galaxy. There is no simple direct relationship between mass and light. 

With this inaccurate assumption, the discrepancy between measured and calculated veloctiy profiles are particularly severe beyond the galactic core. To alleviate this discrepancy, speculations are invoked re ``massive peripheral spherical halos of mysterious Dark Matter'' But no significant matter has been detected in this untenable unstable gravitational halo distribution. This speculated Dark Matter is ``mysterious'' since it does not interact with electromagnetic fields (light) nor ordinary matter except through gravity. This Dark Matter must have other abnormal (non-baryonic) properties to maintain its peripheral spherical shape against the galactic rotation and gravitational attraction of ordinary matter. Many unverified ``mysteries'' are invoked as solutions to real physical phenomena. 

(3) Modified Gravity.  
Re possible deviations from Newtonian gravtiy/dynamics, there is no independent experimental evidence of these deviations. Our use of Newtonian gravity/dynamics with computational techniques has proven successful.   

Conclusion.  
We conclude our approach utilizing Newtonian gravity/dynamics and computationally solving for the ordinary baryonic mass distributions within the galactic disk simulates reality and agrees with data.

\section{Total Galactic Mass}
The measured rotational velocity profiles $V(r)$ (which includes knowledge of $V_{max}$ and $R_g$) are used to compute the galactic rotation parameter $A$. From this information, the total galactic mass
$M_g$ can be calculated via (\ref{eq:parameter-A}) as follows.  
\beq \label{eq:total-mass-Mg}
M_g = \frac{V_{max}^2 R_g}{A \, G} \, . 
\eeq
To check viability we investigate the idealized rotational velocity profile $V(r)$ of our own Milky Way galaxy shown in Fig 1 with $R_c = 0.015$. From data, the parameters appropriate to Milky Way galaxy follow. 

$R_c = 0.015$ 

$V_{max} = 2.5 \times 10^5 (m/s)$ 
  
$R_g = 10^5 (\mbox{light-years}) = 9.46 \times 10^{20} (m)$ 

Our analysis yields the appropriate Milky Way mass distribution (Fig 3) and the galactic rotation parameter $A$ with the following value.  

$A = 1.57$. 

Then from Eq \ref{eq:total-mass-Mg} we find the total galactic mass of the Milky Way as 
 
$M_g = 5.65 \times 10^{41} (kg) = 2.8 \times 10^{11} (\mbox{solar-mass})$  

This value is consistent with Milky Way star counts of 100 billion. Although this is very reasonable agreement, there is additional dust, grains, gases and plasma in our galaxy, so this deduced $M_g$ is on the low side. This deficiency is corrected by employing sphere+disk models as will be discussed in a subsequent companion publication. Sphere+Disk models are more compatible with visual observations of spiral galaxies and yield higher values of $M_g$ more compatible with star counts plus dust, grains, gases and plasma. However, we emphasize the essential physics of galactic rotation is gravitationally controlled by the ordinary baryonic matter within thin galactic disks.  

Incidently note the deduced values of $A$ are typically within a small range around $1.6$ despite an order-of-magnitude variation of the galactic core radius $R_c$.

\section{Limitations and Strengths of Thin Disk Model}
Our simple thin disk model does not address many important features such as spiral structure, plasma effects, galactic formation, galactic evolution, galactic jets, black holes, relativistic effects, galactic clusters, etc.. 

It is well known (Ref.\cite{BT}) that the internal gravitational behavior of a thin disk is much different than a sphere.  This distinctly different behavior enables our thin disk model to describe the rotational dynamics of mature spiral galaxies, and the total galactic mass.  

Our thin disk model has finite radial extent. Beyond the galactic radius, we assume the density has dropped to the inter-galactic level, which is approximately spherically symmetric and thus no longer affects the galactic dynamics. We mention this because some others (Ref.\cite{BT}) have taken the relevant integrals to infinity, which we think is inappropriate. 

In our thin disk approach, we balance the gravitational forces against the centrifugal forces at each and every point. Thus, our solutions for the mass distributions and total galactic mass satisfy the measurements and ensure stability within the same context as similar calculations for our Solar System and Earth satellites. Some previous authors obtain solutions that are not gravitationally stable because they obtain incorrect mass distributions and incorrect galactic masses and do not satisfy the measured rotational profiles. Thus, their solutions are unstable, whereas our solutions are stable within the Newtonian context.  

Plasma effects are certainly active in the formation and evolution of galaxies from the original hot 
plasma (Refs.\cite{Peratt1}-\cite{Peratt3}). However, for the mature spiral galaxies we are addressing, the free plasma density has dropped to levels sufficiently low that plasma does not affect the predominantly gravitational galactic dynamics. This is evidenced in our own Solar System in which gravitational dynamics dominate even in the presence of solar wind, coronal mass ejections, comet tails, etc. The plasma in our Sun is stabilized by gravitational forces, even though plasma effects are very active within the Sun itself. Since our Solar System is approximately 2/3 distance from our Milky Way galactic center, we have our Solar System evidence for the dominance of gravitational forces within our own Milky Way galaxy, at least out to these radial distances.      

Again, we repeat, our thin disk model is sufficient to describe the rotational dynamics of mature spiral galaxies and the total galactic mass.

\section{Applicability of Keplerian Dynamics?}   

In our solar system, the orbital velocities of the planets decrease inversely with the square root of their radial distance $r^{-1/2}$ from the Sun according to the {\em orbital velocity law}   
\[
M_r = r \, V^2(r) / G
\] 
where G is the gravitational constant, and 
$M_r$  
the integrated mass enclosed within a spherical surface of $r$ (called orbital distance).
This Keplerian approximation is based on a simple spherically symmetric gravitational field.  
One can conveniently calculate the (spherically symmetric) mass distribution from
a simple equation with a given rotation curve $V(r)$. 

However, the internal gravitational dynamics of a galactic disk behaves very differently (Eq.2) (Ref.\cite{BT}).
The mass (density) distribution cannot be easily calculated from a simple equation; rather,
it must be computed numerically by solving complicated equations (\ref{eq:force-balance0})
and (\ref{eq:mass-conservation}).  
Moreover, the constraint of mass conservation (\ref{eq:mass-conservation}) enables
the calculation of the galactic rotation parameter $A$ as part of the numerical solution.
Thus, the total galactic mass $M_g$ can be determined from the given $V_{max}$ from the 
measured rotation curve, $R_g$, and the value of $A$.

The experimental fact that galaxies exhibit constant (flat) veloctiy profiles (beyond the core) (Eq.1 and Fig.1), immediately indicates that Keplerian dynamics is not generally applicable within the disk out to the periphery. The reason is that these locations interior to the galactic disk are strongly influenced by the adjacent matter within the disk as we have found. 
Note that it is the total galactic mass $M_g$ and outermost galactic radius $R_g$ that are critical. 
This is another way of expressing the concept that dynamics within the disk are strongly influenced by the adjacent matter within the disk. 
Thus, Keplerian concepts should not be applied to the interior dynamics of a galactic disk.

\section{Conclusions}
The measured rotation velocity profiles of mature spiral galaxies are successfully described with a thin disk gravitational model. Our approach utilizes Newtonian gravity/mechanics to computationally solve for radial mass distributions that satisfy the measured rotational velocity profiles. Our deduced exponential mass distributions in the central core are in agreement with almost all others. However our mass distributions differ out toward the galactic periphery with more mass located in these outer regions which are cooler with lower opactiy/emissivity (and thus darker). Most previous research assumes a galactic density decreasing exponentially with radius out to the galactic periphery, analogous to the  measured light distribution. But this assumption (by others) is inaccurate since both the temperature and opacity/emissity are important but ignored variables. There is no simple relationship between mass and light. These prior models do NOT describe the measured velocity profiles, and speculations are invoked re halos of mysterious Dark Matter or gravitational deviations to compensate. The Dark Matter must have ``mysterious'' (non-baryonic) properties because there is no evidence of its existence and it is not responding to gravitational, centrifugal and electromagnetic forces in any known manner. 
By contrast, our results indicate no massive peripheral spherical halos of mysterious Dark Matter and no deviations from simple gravity. Our total galactic mass determinations are also in reasonable agreement with data.   

The controversy is summarized as follows. 

We believe there is ordinary baryonic matter within the galactic disc distributed more towards the galactic periphery which is cooler with lower opacity/emissivity (and therefore darker). 

Others believe there are massive peripheral spherical halos of mysterious Dark Matter surrounding the galaxies.

\section*{Acknowledgements}
We enthusiastically acknowledge Louis Marmet whose intuition and computational technique convinced us that galactic rotation could be described by suitable mass distributions of ordinary baryonic matter within galactic disks. We also gladly acknowledge Ken Nicholson and Michel Mizony whose similar approaches confirmed our beliefs. Anthony Peratt originally sparked our interest with his plasma dynamical calculations re the formation and evolution of galaxies. Ari Brynjolfsson has energetically supported our efforts.







\begin{figure}[htb]
\resizebox{!}{1.25\textwidth}
{\includegraphics{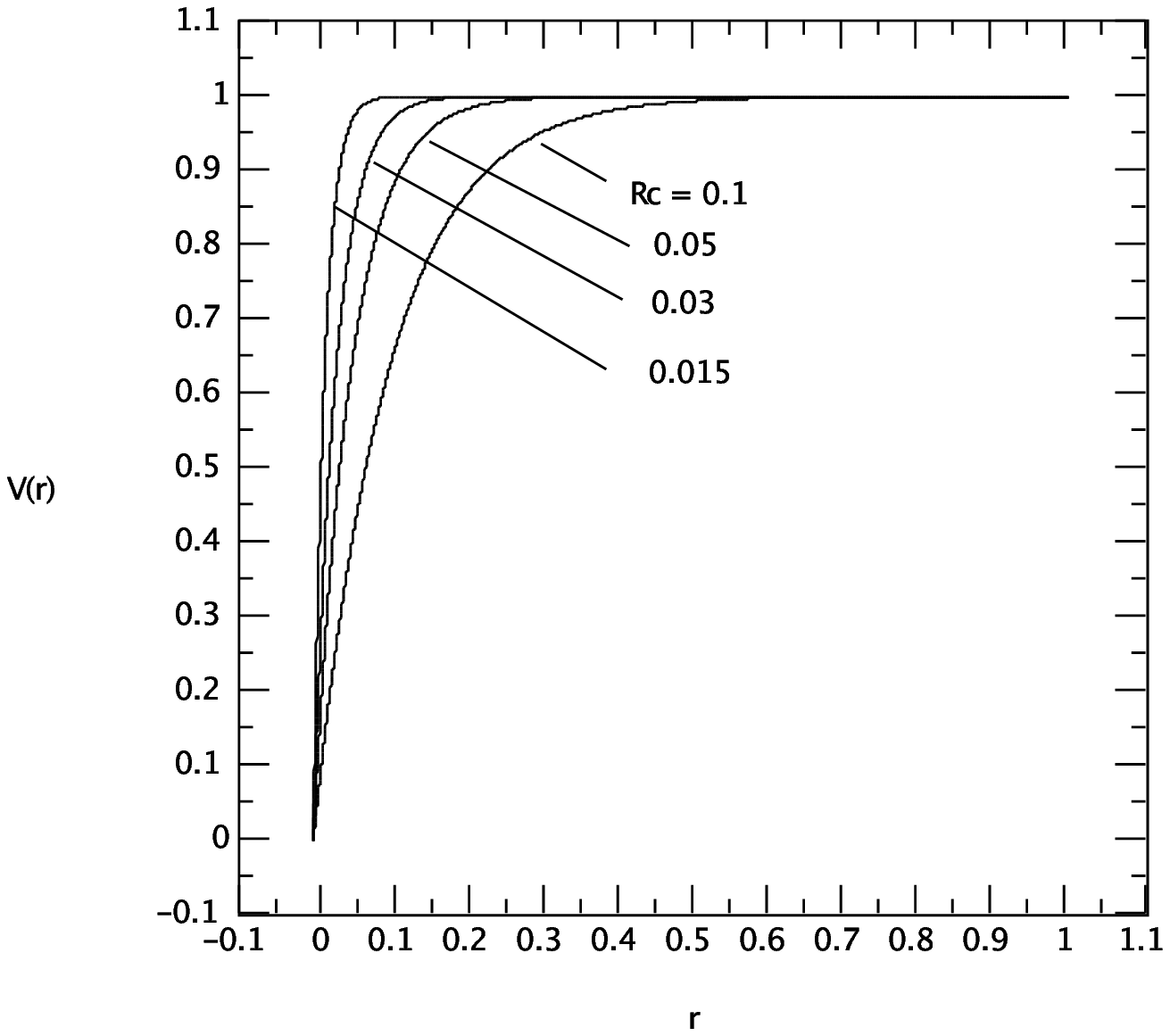}}
\caption{Velocity profiles $V(r)$
according to 
(\ref{eq:smooth-V})
for $R_c = 0.015$, $0.03$, $0.05$ and $0.1$.}
\label{fig:fig1}
\end{figure}

\begin{figure}[htb]
\resizebox{!}{1.25\textwidth}
{\includegraphics{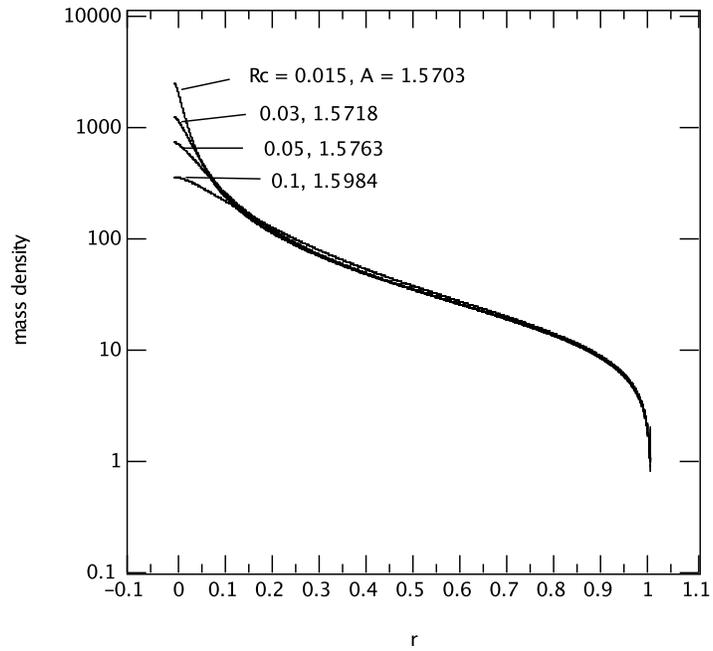}}
\caption{Disk mass density distributions $\rho(r)$ corresponding 
to the
velocity profiles $V(r)$ according to 
(\ref{eq:smooth-V})
for $R_c = 0.015$, $0.03$, $0.05$ and $0.1$
with $A = 1.5703$, $1.5718$, $1.5762$ and $1.5984$.}
\label{fig:fig2}
\end{figure}

\begin{figure}[htb]
\resizebox{!}{1.25\textwidth}
{\includegraphics{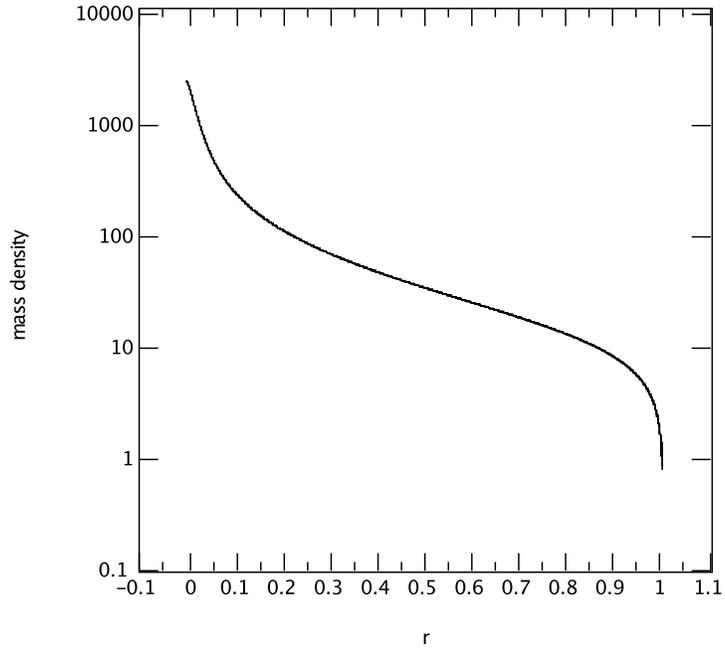}}
\caption{Milky Way mass density distribution $\rho(r)$ corresponding 
to the
velocity profiles $V(r)$ according to 
(\ref{eq:smooth-V})
for $R_c = 0.015$
with $A = 1.5703$.}
\label{fig:fig3}
\end{figure}

\end{document}